\documentclass[letterpaper, aps, showpacs,onecolumn]{revtex4}

\usepackage{amsmath}
\usepackage{amssymb}
\usepackage{latexsym}
\usepackage{graphicx,epstopdf}
\usepackage{hyperref}

\begin{document}

\title{Anomalous quantum-reflection of Bose-Einstein condensates\\ as a self-screening effect}
\author{Alexander Jurisch}
\affiliation{ajurisch@ymail.com, Munich, Germany}
\begin{abstract}
We discuss the effect of anomalous quantum-reflection of Bose-Einstein condensates as a screening effect, that is created by the condensate itself. We derive an effective, time-independent single-mode approach, that allows us to define different paths of reflection. We compare our theory with experimental results.
\end{abstract}
\maketitle

\section{Introduction}
Quantum-reflection is a universal process that occurs when an electrically neutral atom incidences onto a surface with low momentum. The interaction-potential in such an atom-surface system is described by a Casimir-van der Waals-potential tail $V(r)=-\,(\hbar^{2}\beta_{4}^{2}/2m)\,r^{-4}$\,. One remarkable property of Casimir-van der Waals-potential tails is that they generate a quantum-reflection probability $|R_{\mathrm{S}}|^{2}$ that approaches unity when the incident momentum of the particle approaches zero. The universal threshold-law for small momenta $k$ is given by $|R_{\mathrm{S}}|^{2}=1-4\,b\,k+\mathcal{O}(k^{2})$\,, where here $b=\beta_{4}$ is. This behaviour is referred to as normal quantum-reflection, or just quantum-reflection. For an exhaustive review on this topic see \cite{FriTro}\,. However, anomalous quantum-reflection was detected in experiments by Pasquini et. al.  \cite{Pas1,Pas2}\,, where the quantum-reflection of a Bose-Einstein condensate was investigated. In the BEC-case it was found that the quantum-reflection probability does not approach unity for small incident momenta but some smaller value, that appears to be zero for zero incident momenta.

For a linear wave-packet it holds that each single-mode that incidences onto a surface evolves on its own. When the surface is reached, the single-mode eventually becomes reflected or transmitted. By  experiment \cite{Shi, DruDeK, ObeKouShiFujShi, ObeTasShiShi, KouObe}\,, and by theory \cite{CotFriTro, FriJacMei, JurFri}\,, the behaviour of a time-dependent dynamics of a wave-packet matches the behaviour of a single-mode by superposition-effects like a loss of density at the edge of the surface, and by interference. 

For a non-linear wave-packet, however, the global time-dependent decay of the density at the edge of the surface is enhanced by the additional non-linear self-interaction energy. This effect was already assumed in \cite{Pas1, Pas2}\,, demonstrated in \cite{JurRos} and successfully related to the momentum-space in \cite{JurRos1}\,. Thus, it is clear that the additional self-interaction energy must play a crucial role for the understanding of the physics behind the quantum-reflection anomaly.

Scott et. al. \cite{ScoMarFroShe} have performed 3-D simulations of how a condensate behaves when it incidences onto a surface after it is released from a harmonic trap. It was found that the condensate experiences a disruption that goes along with internal excitations. This, however, is to be expected, since the quantum-reflection by the surface generates a momentum-transfer from the surface back into the condensate. By the self-interaction of the condensate this momentum-transfer affects all atoms in the cloud by s-wave scattering. Thus, there exist two currents, one towards the surface and one into the reverse direction. As these currents interact with each other, it may be expected that the atomic cloud should be disrupted.

In the present paper we take a different perspective in order to explain the anomalous quantum-reflection of Bose-Einstein condensates in the single-mode picture. In the case of normal quantum-reflection the single atom, when reflected, just can run away from the surface unhindered. Contrary to this,  in the BEC-case each atom has to move through a fraction of the density-profile in order to reach the surface and, after reflection, has to move through a fraction of the density-profile again in order to be measured. This situation indicates that, on the single-mode level, the effect of screening should play a key-role. 

\section{Self-screening of a Bose-Einstein condensate}
Other than a linear wave-packet, a non-linear wave-packet evolves in the presence of an additional potential that is carried along with the moving wave-packet, because it is generated by the wave-packet itself. In the case of a BEC the self-generated potential emerges from the fact that every single-mode interacts with all other modes that are present in the wave-packet by s-wave scattering. Thus, for each time-step, we can model the behaviour of a single-mode that moves in the presence of all other modes by a time-independent Schr\"odinger equation
\begin{equation}
\frac{2\,m}{\hbar^{2}}\,E\,\phi_{j}({\bf{r}})\,=\,-\nabla^{2}\,\phi_{j}({\bf{r}})\,+\,g\,n({\bf{r}},\,t_{j})\phi_{j}({\bf{r}})\quad,
\label{singlemodeequation}\end{equation}
where the interaction with all other modes is described by the density-potential $n$. The time-evolution of the dynamics of the density-potential $n({\bf{r}},\, t)$ is described by the corresponding Gross-Pitaevskii equation in a fully deterministic way, such that we can construct a map of the interaction-potential $n({\bf{r}},\, t)\,=\,N/V\,|\Psi({\bf{r}},\,t)|^{2}$\,, where the time-evolution can be cut into $J$ single steps. Each of these steps can be treated separately. For the treatment of quantum-reflection we are  interested in the case where the density-potential is located in the very vicinity of the surface. In this situation the condensate is elongated into the perpendicular directions, while in the normal direction the condensate has acquired a higher density, because it is compressed. For quantum-reflection, this justifies to focus on the normal direction. This reduces the problem to one dimension.

The problem that is given by Eq. (\ref{singlemodeequation})\,, with the $x$-axis in normal direction to the surface, is effectively solved by the method of Greens-functions,
\begin{equation}
\phi_{j}(x)\,=\,\exp[\pm\,i\,k_{x}\,x]\,+\,g\,\int\,dx'\,G(x,\,x')\,n(x',\,t_{j})\,\phi_{j}(x')\quad,
\label{solution1D}\end{equation}
where the one-dimensional Greens-function is given by
\begin{equation}
G(x,\,x')\,=\,-\,\frac{i}{2\,k_{x}}\,\exp[i\,k_{x}\,|x\,-\,x'|]\quad.
\label{solution1D0}\end{equation}
Eq. (\ref{solution1D}) is exactly solvable by matrix-inversion. On the lattice Eq. (\ref{solution1D}) transforms into
\begin{equation}
\phi_{j,\mu}\,=\,\exp[\pm\,i\,k_{x}\,x_{\mu}]\,+\,g\,\,G_{\mu\nu}\,n_{\nu}\,\phi_{j,\nu}\quad,
\label{solution1D1}\end{equation}
which leads to
\begin{equation}
\phi_{j,\mu}\,=\,\left(\delta_{\mu\nu}\,-\,g\,G_{\mu\nu}\,n_{\nu}\right)^{-1}\,\exp[\pm\,i\,k_{x}\,x_{\nu}]\quad.
\label{solution1D2}\end{equation}
After the calculation of the single-mode wave-function $\phi_{j}(x)$ according to Eq. (\ref{solution1D}, \ref{solution1D2})\,, we proceed by computing the reflection- and transmission-probablities by standard methods,
\begin{eqnarray}
&&|R_{{\rm{DP}}}(k_{x})|^{2}\,=\,\left|g\,\int_{D}\,dx'\,G(0,\,x')\,n(x',\,t_{j})\,\phi_{j}(x')\right|^{2}\quad,
\nonumber\\
&&|T_{{\rm{DP}}}(k_{x})|^{2}\,=\,1\,-\,|R_{{\rm{DP}}}(k_{x})|^{2}\quad,
\label{amplitudes1D}\end{eqnarray}
where the index DP stands for density-potential. In the following, the reflection-probability at the surface is denoted by 
\begin{equation}
|R_{{\rm{S}}}(k_{x})|^{2}\,=\,\left|\frac{k_{x}\,-\,\sqrt{k_{x}^{2}\,+\,\beta_{4}^{-2}}}{k_{x}\,+\,\sqrt{k_{x}^{2}\,+\,\beta_{2}^{-2}}}\right|^{2}\quad.
\label{surfacereflection}\end{equation}
$|R_{{\rm{S}}}(k_{x})|^{2}$ is the reflection-probability of the step-potential, 
\begin{equation}
V(x)\,=\,-\frac{1}{\beta_{4}^{2}}\,\theta[x\,-\,S]\quad,
\label{steppotential}\end{equation}
where $S$ denotes the position of the surface. As we know that for quantum-reflection the atom-surface potential can be modeled by a step-potential, we can safely use Eq. (\ref{surfacereflection})\,.

We now examine the behaviour of a single-mode by the interplay between the physical presence of the condensate and the surface. We model the normalized condensate-profile in the normal direction and in the vicinity of the surface by a Gaussian
\begin{equation}
|\Psi(x-X)|^{2}\,=\,\left(\frac{1}{\sqrt{\sqrt{\pi\,\alpha^{2}}}}\,\exp\left[-\frac{(x\,-\,X)^{2}}{2\,\alpha^{2}}\right]\right)^{2}\quad.
\label{10}\end{equation}
After the condensate is released from the harmonic trap a Thomas-Fermi shape would certainly apply better, but the compression of the cloud at the surface due to the effect of quantum-reflection makes a Gaussian shape much more likely.

Our assumption is now that a single-mode of the condensate has to tunnel through a fraction of the density-profile in order to reach the surface. At the surface this single-mode then becomes reflected with a certain probability, and thus has to tunnel through a fraction of the density-profile again in order to be measured on the far-side. The fraction of the density-profile can be modeled by $|\Psi(x-X)|^{2}\theta[-(X+\delta)-x]$ for incident single-modes, and by $|\Psi(x-X)|^{2}\theta[x-(X+\delta)]$ for reflected single-modes. The parameter $\delta$ is to be understood as a cut-off parameter. Thereby, we assume that the single-mode always starts to tunnel through the smooth side of the density-profile. This approximately matches the dynamical compression of the condensate immediately before and after quantum-reflection. The joint reflection-probability of the path we have just described is given by
\begin{equation}
\mathcal{R}(k_{x})\,=\,|T_{{\rm{DP}}}(k_{x})|^{2}\,|R_{{\rm{S}}}(k_{x})|^{2}\,|T_{{\rm{DP}}}(k_{x})|^{2}\quad.
\label{firstcase}\end{equation}
Scott et. al. \cite{ScoMarFroShe} have shown that the condensate becomes disrupted by making contact with the surface. Thus, it seems to be likely that single-modes only have to pass through fractions of the density-profile. However, a disintegration is not necessary in the first place, since the described tunnel-effect could also act if the condensate would remain in a coherent state. A disintegration, however, makes it more likely that single-modes only tunnel through fractions of the density before and after contact with the surface.

\section{Comparison with the experiment}
\begin{figure}[t]\centering\vspace{-1.05cm}
\rotatebox{-0.0}{\scalebox{1.2}{\includegraphics{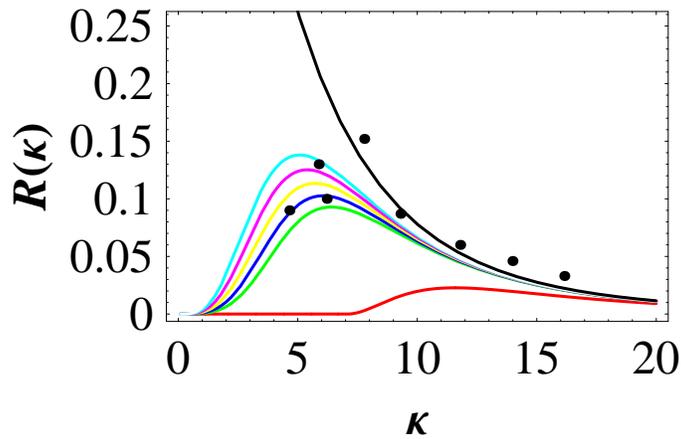}}}
\caption{\footnotesize{Reflection-probabilities in comparison with experimental data from \cite{Pas1}\,, expressed by the black dots. The black curve illustrates normal quantum-reflection. The diverse coloured curves are calculated for different cut-off parameters $\delta$: red ($\delta=0$)\,, green ($\delta=0.15$)\,, blue ($\delta=0.16$)\,, yellow ($\delta=0.17$)\,, purple ($\delta=0.18$) and turquoise ($\delta=0.19$)\,. A match with the experimental results is clearly visible.}}
\label{fig81}\end{figure}
To demonstrate the fidelity of our approach to the anomalous quantum-reflection phenomenon, we now apply it to the situation of the experiment reported in \cite{Pas1}\,. The data of the harmonic trap is given by $(\omega_{x},\,\omega_{y},\,\omega_{z})\,=\,2\,\pi\,(3.3,\,2.5,\,6.5)$ Hz\,, where the $x$-direction is the direction normal to the surface. We convert the frequencies to atomic units and obtain the relevant length-scales of the system, $(L_{x},\,L_{y},\,L_{z})\,=\,(2.19,\, 2.51,\,1.56)\,\times\,10^{5}$ [a.u.]\,. The strength of the atom-surface potential is given by $\beta_{4}\,=\,1.494\,\times\,10^{4}$\,[a.u.]\,, e.g. \cite{JurRos} and references therein. We apply scaling with respect to the direction of normal incidence, such that we obtain $\sigma\,=\,L_{x}/\beta_{4}\,=\,14.66$\,. The quantum-reflection probability by the surface, Eq. (\ref{surfacereflection})\,, thus becomes
\begin{equation}
|R_{{\rm{S}}}(\kappa)|^{2}\,=\,\left|\frac{\kappa\,-\,\sqrt{\kappa^{2}\,+\,\sigma^{2}}}{\kappa\,+\,\sqrt{\kappa^{2}\,+\,\sigma^{2}}}\right|^{2}\quad.
\label{steppotential1}\end{equation}

Now we construct the density of the condensate's interaction-potential by setting $n\,=\,N/V\,|\Psi|^{2}$ as above, and calculate the effective interaction-strength $g_{{\rm{eff}}}\,=\,g\,N/V\,L_{x}^2$\,, where the factor $L_{x}^2$ is due to scaling. The particle-number $N\approx 3\cdot 10^{5}$ is, and from the standard formula for the approximate volume of a condensate, e.g. \cite{PetSmi}, $V\,=\,(L_{x}\,L_{y}\,L_{z})\left(g\,N/(L_{x}\,L_{y}\,L_{z})^{1/3}\right)^{3/5}$\,, we obtain $g_{{\rm{eff}}}\,=\,25.71$\,, where we have used $a_{{\rm{int}}}\,=\,65.3$ [a.u.]\,, e.g. \cite{JurRos} and references therein. The scaled momentum is given by $\kappa=k_{x}\,L_{x}$. For the scaled width of the condensate we have chosen $\alpha/L_{x}=0.31$\,, which leads to a width $\alpha\approx 67,890$ [a.u.] in normal direction. This choice for the width proved to give the best results in comparison with the experimental data. It is interesting to note that this choice coincidences with the ideal width of a condensate for which the diversification of the quantum-reflection behaviour of a condensate due to the strength of the atom-surface potential starts to act. This can be read off the phase-diagram in \cite{JurRos}\,.

In Fig. (\ref{fig81}) we show some measurement-data extracted from \cite{Pas1}\,, as well as the theoretical curves that apply to explain the origin of the events of the measurement. The black dots illustrate the measurement. The reflection-probabilities are calculated according to Eq. (\ref{firstcase})\,. The colours show depleted anomalous reflection-probabilities for certain cut-off parameters $\delta$\,. The red curve illustrates the reflection-probability for a single-mode that tunnels through the half of the density-profile, thus $\delta=0$\,. As it can be deduced, this path is negligible since the depletion is far too strong to be measured on the far-side.
The reflection-curves close to the data are given for the following cut-offs: green ($\delta=0.15$)\,, blue ($\delta=0.16$)\,, yellow ($\delta=0.17$)\,, purple ($\delta=0.18$) and turquoise ($\delta=0.19$)\,. The black curve describes normal quantum-reflection.

It can easily be deduced that our suggestion for a single-mode to tunnel through fractions of the condensate gives a good coincidence with the experimental findings. The good agreement supports our elementary single-mode approach and further confirms the disintegration of the condensate by quantum-reflection by making contact with the surface, as it was found by Scott et. al. \cite{ScoMarFroShe}\,.
\begin{figure}[t]\centering\vspace{-1.05cm}
\rotatebox{-0.0}{\scalebox{1.1}{\includegraphics{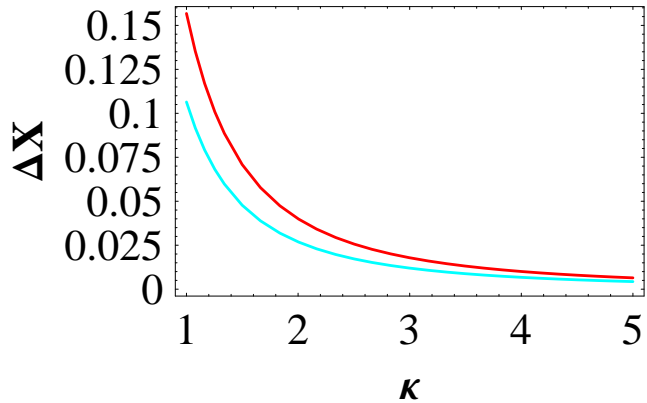}}}
\caption{\footnotesize{Space-shifts $\Delta X$ that are induced by the tunneling-process. The colours match the colours in Fig. (\ref{fig81})\,, red ($\delta=0$)\,, turquoise ($\delta=0.19$)\,.}}
\label{fig82}\end{figure}

To clarify how the tunneling-process influences the time-of-flight of a single-mode we have calculated the induced space-shifts $\Delta X=\Delta x/L_{x}$ by the Eisenbud-Wigner formula. From Fig. (\ref{fig82}) we can read off that for the relevant momenta $\kappa$ a possible delay is already almost zero. Also, as it can be found in \cite{Fri}\,, a time-delay due to the atom-surface potential is negligible on the relevant length-scale $L_{x}$\,.

\section{Conclusion}
The theory we have developed describes the anomalous quantum-reflection of Bose-Einstein condensates as a screening-effect by the fact that a condensate, other than a linear wave-packet where every mode evolves on its own and undisturbed by all other modes, represents a material barrier that can only be traveled through by tunneling.
Our approach is based on the assumption that at a certain time-step we can describe the motion of a single-mode effectively by a one-particle Schr\"odinger equation Eq. (\ref{singlemodeequation})\,, that models the effect of all other modes by a density-potential that is given by an interaction-constant $g$ times the density that is created by all other particles. We also have treated the effect of condensate-screening and quantum-reflection by the surface separately in order to avoid interference between both effects. Interference has not been reported by experiments. Such interference-effects are typical for a time-independent approach, but we have to keep in mind that the reflection-process is of course a time-dependent one, where the screening by the condensate and the reflection by the surface are indeed separated effects on the time-axis. The separation on the time-axis is modeled by our ansatz Eq. (\ref{firstcase})\,, where tunneling through the condensate, reflection by the surface and tunneling into the reverse direction happen subsequently.

We have used the data from the experiment that has been reported in \cite{Pas1}\,. For low momenta, as it is elucidated in Fig. (\ref{fig81})\,, we have found a strong diversification of the reflection-probabilities in the regime where the quantum-reflection probability is depleted. This diversification describes the sensitivity of a single-mode path due to its origin and history during the quantum-reflection process. Also, the diversification matches with the experimental findings \cite{Pas1, Pas2}\,, where the points of measurement close to the saturation maximum of the reflection-probability show some distribution. Generally, our theory is likely to explain the distribution of the measurement-events on an elementary level, by assigning different paths to each event. A measurement-event may stem from a single-mode only, or from a sample of single-modes traveling the same route.

Furthermore, as our approach shows that the self-screening of the condensate takes place only for fractions of the density-profile, the numerical simulations by Scott et. al. \cite{ScoMarFroShe} are supported, where a disruption of the condensate by the action of the quantum-reflection at the surface has been reported.

\subsection{Acknowledgement}
This work is dedicated to my doctoral advisor Prof. Dr. Harald Friedrich, who recently has passed away from a severe disease. Harald Friedrich has dedicated a considerable part of his scientific work to the field of quantum-reflection. He can be regarded as one of the major promoters of the modern theoretical approaches to quantum-reflection phenomena in atom-atom and atom-surface systems.

\end{document}